\documentstyle[wrapft]{ptptex}  
\newcommand{\be}{\begin{equation}}
\newcommand{\ba}{\begin{eqnarray}}
\newcommand{\ee}{\end{equation}}
\newcommand{\ea}{\end{eqnarray}}
\newcommand{\sol}{M_{\odot}}
\newcommand{\soll}{L_{\odot}}

\title{Expected EAGLE event rate towards the Magellanic Clouds}

\author{Takashi {\sc Nakamura}
and Ryoichi {\sc Nishi}$^{*}$}
\inst{Yukawa Institute for Theoretical Physics, Kyoto University,
Kyoto 606-8502, Japan
\\
$^*$Department of Physics, Kyoto University, Kyoto 606-8502, Japan}

\recdate{January 27, 1998
}

\abst{
We  propose to search for MACHOs by observing 
EAGLE (Extremely Amplified Gravitational LEnsing) events
of a majority of dim stars. This search is independent  
of the usual one.
For the  detection limit of EAGLE ($\sim$ 20 mag),  
$\sim 100 ~f ~(\tau^{\rm LMC} / 3 \times 10^{-7})$ 
$(100 ~{\rm days} / \langle \bar t \rangle)$ 
EAGLE events/y are expected to result from all 
the dim stars in LMC. Here $\tau^{\rm LMC}$ and 
$\langle \bar t \rangle$ 
are the optical depth and the average duration of microlensing events, 
respectively, while  $f ~(0 < f < 1)$ is a parameter depending on 
the unknown stellar luminosity function. 
The observed mean duration of EAGLE events  also depends on the 
luminosity function and is $0.01 \sim 0.4$ 
times the usual duration of microlensing events, which corresponds to
 $1 \sim 30$ days. The follow-up observation using larger 
telescopes  may enable us to determine the impact parameter 
and the true duration of the event. If $f$ is determined 
by another independent method, we can 
also determine $\tau^{\rm LMC}$. Even if $f$ is  undetermined, 
the detection of  EAGLE events strongly suggest that MACHOs 
are not due to variable 
source stars, since EAGLE events are due to 
the dim main-sequence stars. 
Although for the SMC, the event rate is smaller 
by a factor of $\sim 7$, it is still a substantial number 
($\sim 13 ~f ~(\tau^{\rm SMC} / 3 \times 10^{-7}$) ($100 ~{\rm days} 
/ \langle \bar t \rangle)$ events/y). 
}

\begin{document}

\maketitle

\section{Introduction}
The analysis of the first 2.1 years of photometry 
of 8.5 million stars in the Large Magellanic 
Cloud (LMC) by the MACHO collaboration\cite{alco96a} 
suggests that the optical depth $\tau^{{\rm LMC}}$ is 
$ 2.9^{+1.4}_{-0.9}\times 10^{-7}$ and 
the mass of MACHOs is $0.5^{+0.3}_{-0.2} \sol$ 
in the standard spherical flat rotation halo 
model.  At present, however,  we do not know where MACHOs are, 
what the fraction of MACHOs in the halo is, and what MACHOs are. 
The basic reason why we do not know the locations and the fraction 
of MACHOs is that the spatial distribution function of MACHOs 
is not known in spite  of many arguments concerning 
the mass distribution of halo dark matter.\cite{pacz86}\tocite{naka96}  
What we can say  from the present observations is  
that the minimum total mass of MACHOs is 
$5.6\times 10^{10} \sol~\tau^{{\rm LMC}}/2\times 10^{-7}$\cite{naka96} 
if the density distribution of 
MACHOs is spherical and a decreasing function of
 the galactcentric radius. This means
that  the  fraction of MACHOs 
up to the LMC is at least 10 \%.

The estimated mass of MACHOs is just the mass of 
red dwarfs. However, the contribution of the halo red dwarfs to MACHO 
events should be small, since the observed density of the halo 
red dwarfs is too low.\cite{bahc94,graf96a,graf96b}
As for the white dwarf MACHOs, 
the mass fraction of white dwarfs in the halo is less than 10 \%, 
since the existence of too many white dwarfs with bright 
progenitors are in conflict with the number counts of distant 
galaxies.\cite{char95,adam96}  
At present we should consider the possibility
that MACHOs may be black holes or boson stars, 
although their origin of existence is completely unknown.

In this situation
the first urgent matter we must consider increasing the statistics of 
MACHO events along the LMC to confirm the optical depth, since 
the total mass of MACHOs becomes smaller than $5 \times 10^{10} \sol$
(i.e. $\sim$ 10 \% of the mass of the halo inside the LMC) 
only if the density distribution of MACHOs is
unusual or $\tau^{{\rm LMC}}\ll 2\times 10^{-7} $.\cite{naka96}   
An independent observation which can confirm the
existence of MACHOs and determine the optical depth is also 
desirable, since MACHO candidates may be variable stars after the 
follow-up 
observation. In reality, the EROS 2 event proved to be 
a variable star.\cite{mils97}

Secondly, from the determination of the optical depth only 
toward the LMC,
for a general non-symmetric density distribution of MACHOs, 
what we can say  is  that the minimum column 
density of MACHOs along the LMC
($\Sigma_{\rm min}^{{\rm MACHO}}$) is 
$25 \sol {\rm pc}^{-2}\tau^{{\rm LMC}} / 2\times 10^{-7}$.
If the clump of MACHOs exists only halfway between LMC and
the sun, $ M_{\rm min}^{{\rm MACHO}}$ is $\sim 1.5\times 10^9 \sol$. 
If this 
is the case, the optical depth in other directions should 
be quite different, and the inhomogeneity of the distribution of 
MACHOs could be checked.\cite{naka96} Therefore to know the spatial 
distribution of MACHOs, 
observation in other directions, such as 
the Small Magellanic Cloud (SMC), is indispensable.

In observations of MACHOs toward the LMC and the SMC, millions of
bright stars with apparent magnitude smaller than $\sim 21$
are observed daily. However, 
there are plenty of stars dimmer than this magnitude limit
(i.e. apparent magnitude larger than $\sim$ 21). 
In this paper we  propose to
search for MACHOs by observing EAGLE events 
(Extremely Amplified Gravitational LEnsing events = microlensing event
with an impact parameter $u \ll 1$) 
of a majority of dim stars. This searchi is independent  
of the usual one. 
In \S 2 we show how to compute the rate and the duration of 
EAGLE events.  In \S 3 we will show the results for the LMC 
and the SMC. \S 4 is devoted to discussion.
 
\section{Event rate of  EAGLE}
EAGLE events in general\cite{wang96} 
and for Bulge sources\cite{goul97} have been discussed. 
Here we discuss EAGLE events 
for the LMC and the SMC. The amplification factor $A$ is given by
\ba
A &=&\frac{u^2+2}{u\sqrt{u^2+4}},\\
u &=&\frac{\theta_s}{\theta_E}, 
\ea
where $\theta_s$ and $\theta_E$ are the angular distance 
to the source  and the angular size of the Einstein radius, 
respectively.
For EAGLE events, $A\simeq 1/u$ is a good approximation.
Now, for a given observation threshold 
of $m_{\rm obs}$ (= $21 \sim 22$ in the apparent magnitude 
for MACHO project), 
if the apparent magnitude ($m_0$) of a star in the LMC (SMC) 
is larger than $m_{\rm obs}$,  
it is so dim that it cannot be identified as a star in 
the LMC (SMC). 
However if it is amplified by a MACHO 
with an impact parameter $u$ smaller than 
$u_{\rm obs} (\equiv 10^{-0.4(m_0 - m_{\rm obs})})$, 
the image of the star can be identified as a star in the LMC (SMC)
 by CCD camera. In practice
we can identify an EAGLE event as a microlensing candidate when 
 the light curve is  well determined so that  
the detection threshold magnitude of the EAGLE event 
$m_{\rm th}$ should be  a little smaller than $m_{\rm obs}$. 
We assume in this paper that
 $m_{\rm th} = m_{\rm obs} - \Delta m$, where $\Delta m ~(\sim 1)$ 
is a constant. 
Then the threshold impact parameter  is given by 
\be
u_T=10^{-0.4(m_0 - m_{th})}.
\ee
We refer to this as an EAGLE event in this paper for definiteness.
EAGLE events will be detected just like new stars, and 
the observational technique is simple. 
Thus, this method is entirely different 
from the so-called pixel lensing method: monitoring the pixels 
rather than the stars.\cite{bai93,goul96,mel97}

The event rate of EAGLE for a source star is given by\cite{grie91}
\ba
\Gamma  &=& \frac{\tau}{\langle \bar t \rangle} u_T, \\
          &\simeq& 1.1 \times 10^{-6} 
          \left( \frac{\tau}{3 \times 10^{-7}} \right)
          \left( \frac{\langle \bar t \rangle} 
            {100 ~{\rm days}} \right)^{-1} ~u_T ~~
           {\rm events/year}.
\ea
where $\tau$ and $\langle \bar t \rangle$ are the optical depth 
and the average duration ($u \leq 1$) of microlensing events, 
respectively.
Note that $\Gamma$ is proportional to the luminosity 
of the source star.
With $n_L(m_0)$ and $m_T$ being the stellar 
luminosity function and the total apparent magnitude of the LMC (SMC), 
respectively, we have  the total event rate ($\Gamma_E$) of EAGLE 
by all  stars in the LMC (SMC)  as 
\ba
\Gamma_E &=& \int^\infty_{m_{\rm obs}} \Gamma n_L(m_0) dm_0, \\
         &\simeq& 1.1 \times 10^{-6} ~f ~10^{-0.4(m_T - m_{\rm th})}
         \left( \frac{\tau}{3 \times 10^{-7}} \right)
         \left( \frac{\langle \bar t \rangle}{100 ~{\rm days}} 
                        \right)^{-1} 
                 {\rm events/year}, \\
  &\sim& 100 ~f ~\left( \frac{\tau}{3 \times 10^{-7}} \right)
  \left( \frac{\langle \bar t \rangle}
    {100 ~{\rm days}} \right)^{-1} {\rm events/year}  
  ~{\rm for ~LMC} ~(m_T=0.1), \\
  &\sim& 13 ~f ~(\frac{\tau}{3 \times 10^{-7}})
  \left( \frac{\langle \bar t \rangle}
    {100 ~{\rm days}} \right)^{-1} {\rm events/year}  
  ~{\rm for ~SMC} 
  ~(m_T=2.3), 
\ea
where $f$ is the  fraction of the luminosity by the stars below  the
observation threshold ($m_0>m_{\rm obs}$) to the total luminosity of 
the LMC (SMC). We assume $m_{\rm th}=20$.

Since the surface brightness of the LMC and the SMC are not so large, 
 even for the central part of these galaxies
 the surface brightnesses are B(0) = 21.17 mag ${\rm arcsec^{-2}}$ 
for the LMC 
and B(0) = 22.71 mag ${\rm arcsec^{-2}}$ for the SMC.\cite{both88} 
 This means in ordinary fields $m_{\rm obs}$ is $\sim$ 21 mag.
In reality the MACHO project monitors many stars 
of magnitude $\sim$ 21 mag.
Thus, the magnitudes for the EAGLE events can be detected similar to 
the usual MACHO events. 

Since there is no available luminosity function for the LMC 
and the SMC 
except for fairly brighter stars, we estimate $f$ using the following  
two methods; 

Model 1: We use the luminosity function for our Galaxy to estimate 
$f$.

Model 2: We assume a power law stellar initial mass function 
with the power index $\alpha$ and a constant star formation rate.

\subsection{Model 1}
The luminosity function of Scalo\cite{scalo86} 
can be approximated as  
\ba
\log \phi(V) &=& \frac{3}{5}V - 4.2  ~~(V \leq 2) \\
             &=& \frac{1}{8}V - 3.25  ~~(2 \leq V \leq 10) \\
             &=& -2  ~~(10 \leq V), 
\ea
where $V$ is the stellar absolute visual magnitude. 
Then, we have 
\be
f = \frac{\int_{m_{\rm obs}-\mu}^\infty L_0(V) \phi (V) dV}
{\int_{-\infty}^\infty L_0(V) \phi (V) dV}, 
\ee
where $\mu$ is the distance modulus and $L_0$ is the absolute 
luminosity
of the star.  The results are shown in Tables I to IV as $G$.

\subsection{Model 2}
We adopt the power law initial mass function as
\be
n(M)  =  CM^{-\alpha}, ~~~~(M_l < M < M_u)
\ee
where $M_l$ and $M_u$ are 
the lower and the upper mass limit of the stars, respectively.
The mass-luminosity relation of the main sequence star is expressed 
by\cite{kipp90}
\ba
L_0(M)&=&\soll \left( \frac{M}{\sol} \right)^\eta, \\
\eta &\sim& 3, 
\ea
where $M$ is the mass of the star.
The observation of the color distribution suggests 
that neither the LMC nor the SMC is likely experiencing 
a star-formation 
``burst'' at the present epoch, both galaxies have formed 
a bulk of their stars for the last 5 Gyr,\cite{both88,feast95} 
and the star formation rates are 
fairly constant.\cite{rocca81,gall96} 
Thus, for simplicity, we assume that the star formation rates of 
these galaxies have been  
constant for the last 5 Gyr and was zero before 5 Gyr ago.
Assuming that the lifetime of the sun is 10 Gyr, 
we obtain the present day mass function as
\ba
n_p(M)  &\simeq& CM^{-\alpha} ~~(M < 1.4 \sol ), \\
        &\simeq&  2 \left( \frac{M}{\sol} \right)^{-2} CM^{-\alpha} 
                ~~(1.4 \sol \leq M).
\ea
Initial mass functions of massive stars 
of the LMC and the SMC are not so different from that of 
the Galaxy,\cite{mass95} 
and the slopes for OB associations 
are found to be essentially the same as Salpeter IMF.\cite{sal55}
However, for field stars, the slope is very steep as $\alpha \sim 5$.
Taking account of these data, we assume 
a single power law and use the two extreme values of 
$\alpha = 2.35$ and 5 to know the range of $f$ which is given by

\be
f = \frac{\int_{M_l}^{M_{\rm obs}} M^\eta n_p(M) dM}
{\int_{M_l}^{M_u} M^\eta n_p(M) dM}, 
\ee
where $M_{\rm obs}$ is the mass of the star whose apparent magnitude 
is $m_{\rm obs}$.

We next discuss the duration of an EAGLE event ($t_E$).
In a microlensing event, the apparent luminosity as 
a function of time ($l(t)$) is expressed as
\be
l(t)=\frac{R_E l_0}{\sqrt{b^2+v^2_\perp(t-t_0)^2}}, 
\ee
where $R_E, b, v_\perp$ and  $t_0$ are the Einstein radius,
 the impact parameter, the transverse velocity 
and the time of the maximum luminosity, respectively,  
and $l_0 = 10^{-0.4 \mu} ~L_0$. From the observational data, 
we can determine the values of $t_0$ and 
$R_E l_0/b$. By a follow-up observation using large telescopes
we may determine $l_0$ so that 
$R_E/b$ $=u_{\rm min}^{-1}$ may be determined.
$t_E$ is defined to be the duration for which the apparent magnitude 
($m$) of the source star satisfies $m < m_{\rm obs}$. 
This quantity is given by
\ba
t_E &=& t_{\rm dur}\sqrt{\frac{l_0^2}{l_{\rm obs}^2}-
        \frac{b^2}{R_E^2}}, \\
    &=& t_{\rm dur}\sqrt{u_{\rm obs}^2-u_{\rm min}^2}, \\
    t_{\rm dur} &=& \frac{2R_E}{v_\perp},
\ea 
where $l_{\rm obs}$ is the apparent luminosity corresponding to 
the observation threshold magnitude $m_{\rm obs}$ and 
$u_{\rm obs} = 10^{-0.4 (m_0 - m_{\rm th})}$. 
From the observed values of 
$t_E, m_0$ and $b/R_E $, we can determine 
the true duration of microlensing event $t_{\rm dur}$.
The mean value of $t_E(m_0)$ for given $u_T(m_0)$ is derived as
\ba
\langle t_E(m_0) \rangle &=& t_{\rm dur}\frac{1}{u_T}\int^{u_T}_0 
\sqrt{u_{\rm obs}^2-u_{\rm min}^2} du_{\rm min}, \\
  &\simeq& 0.97 t_{\rm dur}u_{\rm obs}(m_0), 
\ea 
where we assume $\Delta m =1$.
The mean value of $\overline{\langle t_E(m_0) \rangle} $ 
for different $m_0$ is given by
\be
\overline{\langle t_E(m_0) \rangle}=\frac{1}{\Gamma_E}
\int^{m_{\rm th}}_{\infty} \langle t_E(m_0) \rangle 
\Gamma n_L(m_0) dm_0.
\ee

\section{Results}

We computed $f$ and $\overline{\langle t_E(m_0) \rangle}$ 
for the LMC and the SMC 
using the Galactic stellar luminosity function (marked as $G$) 
as well as  using the power law stellar model with $\eta=3$, $M_u=50
 \sol$, $M_l=0.1 \sol$, 
and the power index $\alpha = 2.35$ and 5 for two cases of 
the detection threshold as

Case A) $~m_{\rm obs}=21$ mag, $m_{\rm th}$ = 20 mag 
and $\Delta m$ = 1 mag.

Case B) $~m_{\rm obs}=19$ mag, $m_{\rm th}$ = 18 mag 
and $\Delta m$ = 1 mag.

\noindent
As for $\mu$ and $m_T(V)$,  we adopt $\mu=18.5$ mag 
and $m_T(V)=0.1$ mag 
for the LMC and $\mu=18.9$ mag and $m_T(V)=2.3$ mag for the SMC.

\begin{wraptable}{r}{\halftext}
%\begin{table}
\caption{Case A for LMC}
\label{table:1}
\begin{center}
\begin{tabular}{cccc} \hline \hline 
$\alpha$ & $f$ & $\overline{\langle t_E(m) \rangle}/t_{\rm dur}$ 
\\ \hline
2.35 & 0.36  & 0.28 \\
5.0 & 0.99 & 0.02 \\ \hline
$G$ & 0.32 & 0.41 \\ \hline
\end{tabular}
\end{center}

\vskip 0.6cm

\caption{Case A for SMC}
\label{table:2}
\begin{center}
\begin{tabular}{cccc} \hline \hline
$\alpha$ & $f$ & $\overline{\langle t_E(m) \rangle}/t_{\rm dur}$ 
\\ \hline
2.35 & 0.40 & 0.26 \\
5.0 & 0.99 & 0.02 \\ \hline
$G$ & 0.41 & 0.41 \\ \hline
\end{tabular}
\end{center}

\vskip 0.6cm

\caption{Case B for LMC}
\label{table:3}
\begin{center}
\begin{tabular}{cccc} \hline \hline
$\alpha$ & $f$ & $\overline{\langle t_E(m) \rangle}/t_{\rm dur}$ 
\\ \hline
2.35 & 0.54 & 0.18 \\
5.0 & 1.00 & $5 \times 10^{-3}$ \\ \hline
$G$ & 0.72 & 0.26 \\ \hline
\end{tabular}
\end{center}

\vskip 0.6cm

\caption{Case B for SMC}
\label{table:4}
\begin{center}
\begin{tabular}{cccc} \hline \hline
$\alpha$ & $f$ & $\overline{\langle t_E(m) \rangle}/t_{\rm dur}$ 
\\ \hline
2.35 & 0.58 & 0.16 \\
5.0 & 1.00 & $4 \times 10^{-3}$ \\ \hline
$G$ & 0.77 & 0.22 \\ \hline
\end{tabular}
\end{center}
%\end{table}
\end{wraptable}

The results are shown in Tables I to IV.  Using these tables,
we display the expected number of EAGLE events for all MACHO
spherical halo in the standard flat rotation curve 
($\tau^{\rm LMC}\sim 5\times 10^{-7}$ and 
$\tau^{\rm SMC}\sim 7\times 10^{-7}$)\cite{grie91} 
with $\langle \bar t \rangle$ = 70 days since all MACHO halo is still
possible from the observational data.\cite{alco96a} Then, for Case A,  
we have $\Gamma_E^{\rm LMC}=240 ~f$ events/year 
and $\Gamma_E^{\rm SMC}=31 ~f$ 
events/year, so that even for the smallest value of $f$, we expect 
$\Gamma_E^{\rm LMC}=77$ 
events/year and $\Gamma_E^{\rm SMC}=13$ events/year.
For case B, with the smallest value of $f$, we expect 
$\Gamma_E^{\rm LMC}=21$ events/year 
and $\Gamma_E^{\rm SMC}=3$ events/year.

Are there possible EAGLE events in already existing data? 
This is certainly possible. One
example is LMC event 12 of the MACHO project, which may be 
an EAGLE event.  
\cite{alco96a} Thus it is necessary to 
 search EAGLE events systematically and estimate the detection 
efficiency even using existing observational data. 
However, the duration of EAGLE events is usually short 
(1 day $\sim$ 30 days) 
especially for Case B, so that
 the usual observational mode (1 or 2 observation/night)
is not adequate. In this sense the observational mode used by the MOA
collaboration (Japan-NewZealand collaboration of 
Microlensing Observation
for Astronomy,\cite{abe96} for example, seems to be suitable.
There working on this project have been trying to observe 
microlensing events by planet MACHOs
since May 1996 at Mt. John, so that they have been observing 
 stars in the LMC and the SMC
as frequently as possible. This year they are planning to observe
1.5 million stars in the LMC 12 times/night in the winter 
and 6 times/night in the summer.\cite{mura97} 
If 8 microlensing events reported by
the MACHO collaboration are not due to variable source stars 
but due to
MACHOs, a substantial number of EAGLE events should be observed 
in the observation mode taken such as by the MOA collaboration.  

\section{Discussions}
In the actual observation of microlensing events, only a fraction of 
the LMC and/or the SMC are observed. 
The LMC and the SMC are too large to be  monitored
frequently since 
half light radii are estimated to be $3.03 ~\pm ~0.05$ degree (LMC) 
and $0.99 ~\pm ~0.03$ degree (SMC),\cite{both88} respectively. 
 This means that the detection rate of EAGLE should be multiplied by 
$g\equiv$ (luminosity of the observed area)/(the luminosity of LMC). 
Although for the LMC,  $g$ is small  ($g\sim 0.2$), the
event rate of EAGLE is substantial ($\sim$ 20 events/year). 
 For the SMC 
$g$ can be large ($g\sim 0.5$)  since the SMC is smaller than the LMC, 
 so that the event rate of EAGLE 
is also substantial ($\sim$ 10 events/year). 

{}From the observational event rate of EAGLE, we know the quantity
$f\tau$  from Eqs. (8) and (9).
To estimate $f$ we used the stellar luminosity function of the Galaxy 
as marked $G$, and Salpeter IMF 
($\alpha =2.35$) to obtain similar results. 
If the star formation history and/or stellar 
initial mass functions of the LMC and the SMC are different from 
those of the Galaxy, the luminosity functions should be different from 
the Galaxy. 
However, the initial mass functions of massive stars 
of these galaxies are fairly similar to 
that of the Galaxy,\cite{mass95} and the current star formation is not
very active,\cite{both88} so that  the deviation of $f$ from 
that of case $G$ may not be large.  
This is a completely independent determination of 
$\tau$ from the usual method.

For EAGLE events we observe extremely amplified dim stars, 
which are mostly main sequence stars. 
Since the intrinsic variabilities of these stars 
are expected to be small, we can pick up microlensing events 
efficiently. 
Moreover, the EAGLE event rate toward the LMC estimated in this paper, 
$\sim $ 20 events/year, may be much larger 
than the usual non-EAGLE MACHO event rate, $\sim 4$ events/year. 
If $\tau^{\rm LMC} \sim 3 \times 10^{-7}$, 
numerous EAGLE events must be observed toward the LMC so that
the observation of EAGLE events will give us an independent method to 
confirm the existence of MACHOs.
As for the SMC, the EAGLE event rate  is also substantial, and
 the observation of EAGLE events
toward the SMC may determine the optical depth toward the SMC, 
which is needed
to know the spatial distribution of MACHOs. 

\section*{Acknowledgments}
We would like to thank Y. Muraki for useful comments.
 This work was supported by the 
Grant-in-Aid for Scientific Research from the Ministry of Education,
Science, Sports, and Culture, No. 09640351 (TN),  No. 09740174 (RN).

\end{document}